# MEAN-MOTION RESONANCES IN THE QUADRANTID METEOROID STREAM AND DYNAMIC EVOLUTION OF DUST TRAIL OF ASTEROID (196256) 2003 EH1

**G. E. Sambarov[1], A. P. Kartashova[2], and T. Yu. Galushina[1]**

UDC: 521.1, 521.182

*The paper presents numerical models of mean-motion resonances detected in the Quadrantid meteoroid stream consisting of particles of mass 0.003 to 0.03 g, which helps to prove the presence of the following mean-motion resonances: the 1:9 resonance with Venus, 1:5 resonance with Earth, 1:3 and 3:8 resonances with Mars, and 2:1, 7:3, 9:4 resonances with Jupiter. Resonance particles create dust trails far from the Earth's orbit. Indeed, there is no observational support for resonant effects in the observed Quadrantid meteoroid stream.*

**Keywords:** numerical simulation, asteroid 2003 EH1, Quadrantid meteoroid stream, mean-motion resonance.

## INTRODUCTION

This work studies the evolution of the model Quadrantid meteoroid stream released from the near-Earth asteroid (196256) 2003 EH1 [1, 2]. The Quadrantids is considered to be one of the best annual meteor showers, which peaks on January 3–4. The first observation of the Quadrantid shower was made in 1835, although some observations go back to 1798 [1]. A rapid change in the nodal distance of the Quadrantid meteoroid stream or its younger age (the Earth's orbit started to pass through this stream recently) leads to its observations since recently.

There is still disagreement in the world community as to a single model of meteor streams. Therefore, each meteor stream requires an individual scientific approach. The age and formation mechanism of the core of the Quadrantid meteoroid stream, are investigated in [1, 2]. According to Kasuga and Jewitt [3], the "current dust production from 2003 EH1 is orders of magnitude too small to supply the mass of meteoroid stream in the 200–500 year dynamical lifetime. If 2003 EH1 is a source of the Quadrantids, we infer that mass must be delivered episodically, not in steady-state."

The evolution of the mean orbit of the Quadrantid stream and its meteoroids has been studied many times. Hughes *et al.* [4] show that the node motion is very sensitive to the orbit parameters used, whereas Williams and Wu [5] demonstrate that the mean orbit undergoes multiple close approaches to Jupiter. This means that meteoroids initially close to each other in phase-space orbits, in a short time can separate from each other to a very long distance. Numerical studies [6, 7] of the motion of 2003 EH1 and its 500 clones show a rapid departure of the latter from the nominal orbit due to multiple close approaches to planets. This is consistent with all stated above. Thus, in the case of the Quadrantid meteoroid stream, any initial structure will be lost very rapidly.

The aim of this work is to study mean-motion resonances in the Quadrantid meteoroid stream and propose a scenario for the dynamics of the 2003 EH1 dust flux. The focus of this work is on the probable production of dust particles or clusters in space.

## 1. MEAN-MOTION RESONANCES IN METEOROID MOTION

The numerical simulation was used to investigate the mean-motion resonance (MMR) effect on the particle motion in the Quadrantid meteoroid stream. It included the following stages:

– the probabilistic evolution of the orbit was numerically simulated with the IDA software package (Investigating Dynamics of Asteroids) [8];

– identification of close approaches of the meteoroid stream to planets;

– application of Mean Exponential Growth factor of Nearby Orbits (MEGNO), a tool to investigate the dynamic behavior of space objects [9, 10];

– numerical and analytical method for the MMR identification.

The dynamic behavior of meteoroid particles in the Quadrantids was investigated by the numerical integration of equations of motion by using Everhart's method [11, 12] in the IDA [9]. At the stage of the preliminary MMR identification, we simulated particles splitting from the Quadrantid parent body. The model of forces included perturbations of major planets, Moon, Pluto, Ceres, Pallas, Vesta as well as the light pressure, relativistic effects from Sun, and the Poynting-Robertson effect. The resonant motion was obtained *via* the analysis of two parameters, namely resonant (critical) argument β

$$\beta = k_1 \lambda_1 - k_1 \lambda_2 - (k_1 - k_2)\omega_1 - (k_1 - k_2)\Omega_1 \qquad (1)$$

and the so-called resonant gap α [13], which is a first time derivative of β:

$$\alpha \approx k_1 n_1 - k_2 n_2, \qquad (2)$$

where ω is the perihelion argument; Ω is the longitude of ascending node of the asteroid orbit; $n_1$, $n_2$ are the mean motions of asteroid and planet, respectively; $\lambda_1$, $\lambda_2$ are mean longitudes of asteroid and planet, respectively; $k_1$, $k_2$ are integers.

---

[1] National Research Tomsk State University, Tomsk, Russia, e-mail: detovelli@mail.tsu.ru; Tatyana.Galushina@mail.tsu.ru;
[2] Institute of Astronomy of the Russian Academy of Sciences, Moscow, Russia, e-mail: akartashova@inasan.ru

Resonance occurs when the gap approaches to zero, while the argument makes periodic oscillations within 360 degrees. When the resonant gap α strictly equals zero, it is called a precise commensurability or precise resonance. The similar algorithm of the MMR calculation is described in [8, 13].

Equations of motion combined with equations of the MEGNO parameters written in the inertial system, are numerically integrated by the high-order Gauss–Everhart integrator [12]. In all cases, coordinates of major planets, Pluto and the Moon are determined by fundamental ephemeris DE431 based on the Jet Propulsion Laboratory Development Ephemeris [14]. Relativistic effects are considered by equations of the relativistic motion of Schwarzschild.

## 2. SIMULATION OF THE QUADRANTID METEOROID STREAM

The simulation approach is well developed by many researchers [1, 2, 4, 15]. Today, it is found that meteoroid streams mostly appear after disintegration of cometary nuclei. All models of the meteoroid release considerably unify the same physical concepts with slight changes. The main idea of this approach is the release of a certain number of test particles at certain points on the orbit of the parent body and control for the behavior of each of them in the selected period. The current particle release from 2003 EH1 is too small [3] to provide the mass of meteoroid stream between 200 and 500 years ago. Numerical integration of equations of motion of 2003 EH1 and its 500 clones [6, 7] shows that clone orbits rapidly move away from the nominal orbit due to multiple close approaches to planets.

The models of dust ejection from cometary nuclei suggest that the released mass consists of gas and dust, the latter having no high velocity (1–100 m/s) as compared to the cometary nucleus. The ejection velocity is much slower (a few dozens of meters per second) than the cometary orbital velocity (a few dozens of kilometers per second in perihelion). The lowest relative velocities of 200 to 800 m/s are given by Abedin [1] for simulated particles of the core of the Quadrantid meteoroid stream from 2003 EH1, most of its clones having relative velocities exceeding 1 km/s of characteristic particle velocity. In modeling, the dust ejection velocity within 200–800 m/s will be a reasonable compromise.

An artificial model, not describing the real properties of either the stream, or ejection, can be used for this purpose. We simulated the ejection of 3000 test particles of the Quadrantids on the perihelion orbit of 2003 EH1. The particle ejection velocity was the same and calculated from the Whipple formula [15]. The osculating orbit of the asteroid was used as a reference for the period of August 10, 1769 (JD = 2367396.5) for the perihelion passage. The particle release was simulated as isotropic. In this work, we considered three model subflows having different masses. Each subflow consisted of 1000 particles with masses $m_1 = 0.3$ g, $m_2 = 0.03$ g, $m_3 = 0.003$ g and density of 1 g·cm$^{-1}$. Isotropic release in the perihelion orbit of the parent body was integrated backwardly to the initial model period.

## 3. ORBIT EVOLUTION OF THE QUADRANTID METEOROID STREAM

This section presents the analysis of the probable orbital evolution of the Quadrantid meteoroid stream based on the motion of test particles released from the parent body. The orbital evolution of each ejected particle is studied till the year 4000. Research findings are presented in Fig. 1. One can see the longitude of ascending node Ω, argument of perihelion ω, semi-major axis *a*, orbital inclination *i* to ecliptic plane, and eccentricity *e*.

In Fig. 1, the test particle behavior is presented in gray color, while the asteroid orbit is black. The semi-major axis *a* is relatively stable and approaches to 3.1 a.u. It means that the energy of the mean orbit of the stream remains constant during the whole period regardless of a series of close approaches to planets.

According to simulation results of the orbit of meteoroids released from the parent body in the past, on average just under half of ejected particles move chaotically [7]. The motion of these objects over long time scales, is considered to be chaotic due to probably repeated close approaches of released particles to Jupiter and their motion nearby the 2:1 MMR with Jupiter causing "weak chaos" [2]. This "weak chaos" is largely confined to the true anomaly. The orbit shape can therefore be computed reliably over much longer time scales than can the body's position within the orbit.

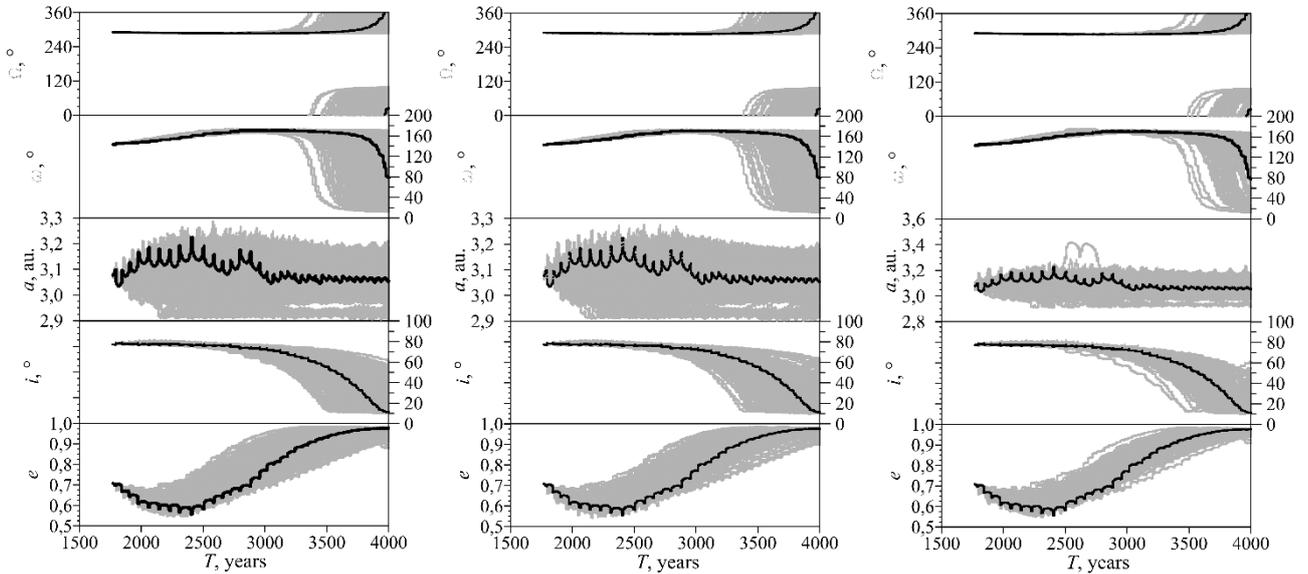

Fig. 1. Probable orbital evolution of test particles (gray color) and 2003 EH1 (black color) for the period from 1769 AD till 2019 AD. Left to right: $m_1$, $m_2$ and $m_3$ test particles.

The MEGNO ($\bar{Y}(t)$ parameter) allows distinguishing between chaotic and regular motions and measure the time of their predictability [16]. The temporal evolution of $\bar{Y}(t)$ demonstrates characteristics typical for various orbits. The $\bar{Y}(t)$ parameter growth is linear for the chaotic orbit, while for the regular orbit it is lower or about 2. For example, for quasi-periodic (regular) orbits, $\bar{Y}(t)$ always tends to 2 and is zero for stable orbits of the harmonic oscillator type. The $\bar{Y}(t)$ parameter grows for test particles of the meteor stream moving nearby the Sun and massive planets. High MEGNO values are conditioned by frequent changes in the semi-major axis caused by multiple approaches to Jupiter nearby the Hill sphere. Note that the chaotic behavior of test particles can be caused not only by close approaches to planets, but also unstable resonances, which will be considered in the next section. Figure 2 presents four cases to refer to illustrate the MMR influence on these particles in sections below. All these cases also relate to other particles ejected from 2003 EH1.

According to Fig. 3, the orbital evolution of 2003 EH1 shows its significant change by the end of the period considered. This is because the approach to Jupiter, overlapping of different apsidal and nodal resonances [7], and the proximity to low-order MMR. An unstable asteroid motion is induced by repeated approaches to Jupiter and unstable MMR with Venus, Mars, and Jupiter as a whole. The resonance overlap leads to chaotic motion of the asteroid.

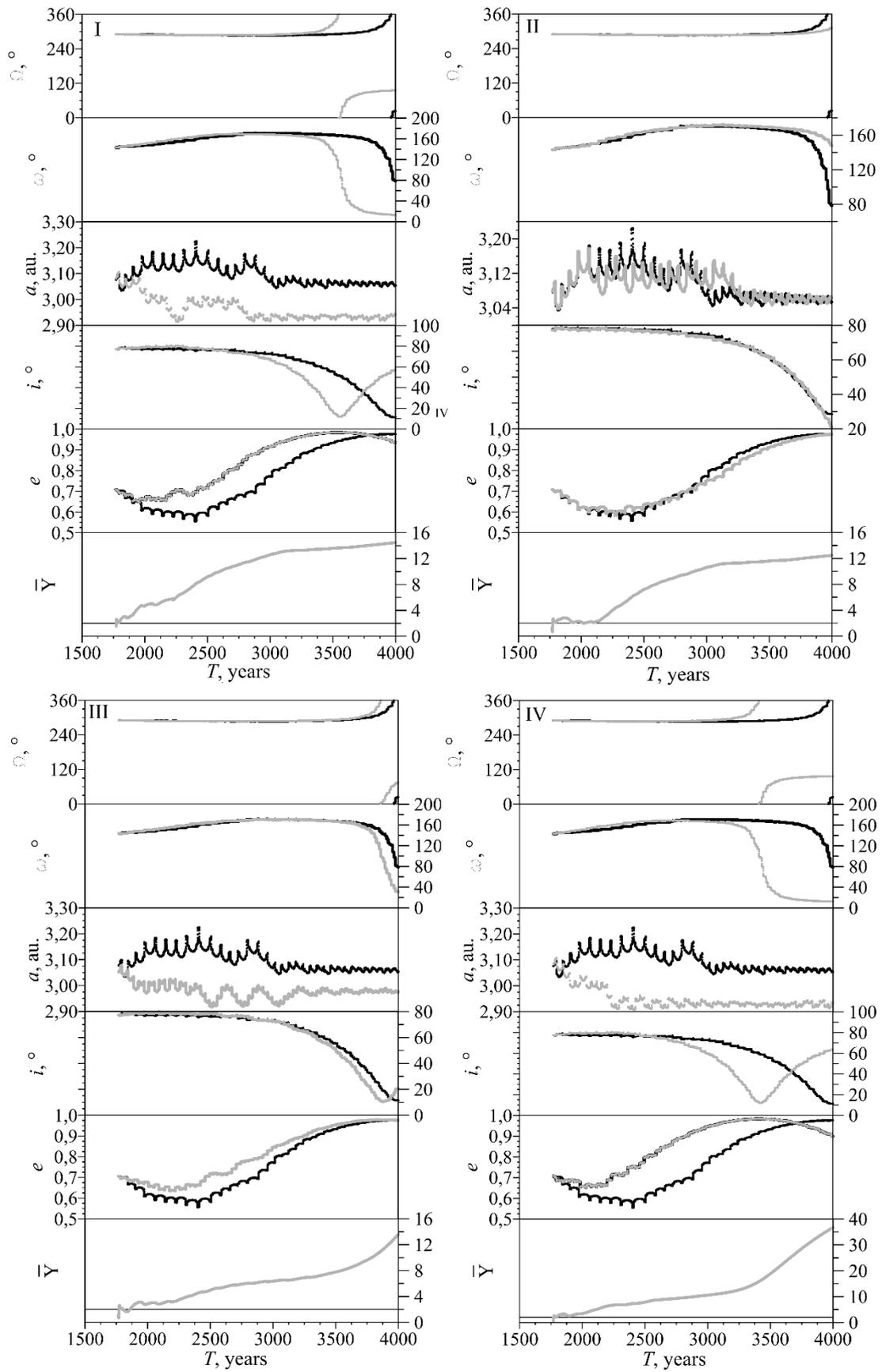

Fig. 2. Probable orbital evolution of test particles (gray color) of the Quadrantids and 2003 EH1 (black color) for the period from 1769 AD till 2019 AD: I, II – $m_1$ test particles, III – $m_2$ test particle, IV – $m_3$ test particle.

As can be seen in Fig. 3, meteoroids initially close to each other in the phase space, in the future move away from each other to a large distance during short period, i.e., space density function of objects changes. The stream takes the form of a torus, its structure is distinctive. In Fig. 3*b*, one can see the same state after 600 years of evolution, i.e., the particles remain part of the stream orbit and acquire a structure typical for a toroidal meteor cluster. After the year 3000, the stream looks quite different: there is no any density growth along the perimeter and no hollow tube. The observed random chaotic effects cause the particle deviation from the orbit plane of the parent body and their ejection from the torus structure of the meteoroid cluster. It is thus can be concluded that in 600 years of evolution, the stream has lost its initial structure, the particles are characterized by the reduced space density, since objects leave the considered region due to perturbations. As we believe in [7], perturbations from inner planets and Jupiter predominate in the formation of the observed structure of the stream. Numerical integration of the motion equations of test particles shows, that in the future, their orbits move away from the nominal orbit of the parent body due to multiple close approaches to planets, which confirms the above mentioned. Hence, in the case with the Quadrantid stream model, any initial structure will be rapidly lost.

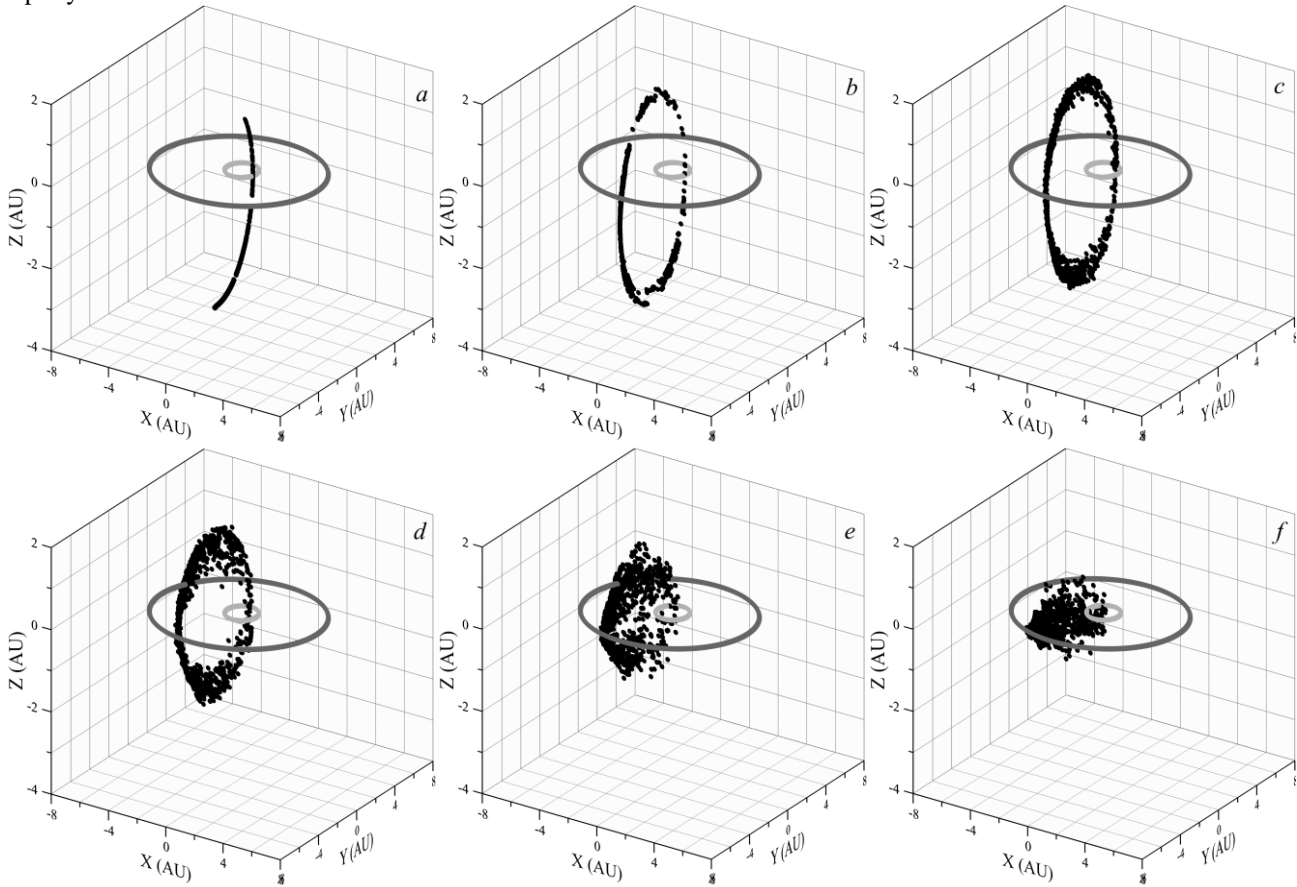

Fig. 3. Three-dimensional simulation of $m_1$ test particles of the Quadrantids (black spots): *a* – December 1834, *b* – January 2000, *c* – January 2500, *d* – January 3000, *e* – January 3500, *f* – January 4000. Gray and light gray colors indicate orbits of Jupiter and Earth, respectively.

## 4. RESULTS AND DISCUSSION

The analysis of the meteoroid particle dynamics hypothetically released from 2003 EH1 approaching to the Earth, shows a complex structure of the Quadrantid meteoroid stream approaching to orbits of Jupiter and Earth. Meteoroids inherit the dynamic properties of the asteroid, but not all of them. Table 1 summarizes mean-motion resonances. A certain number of resonant particles is found for the comet model. It is worth noting that we look for particles presenting just in the resonance region for a long time (over 100 years).

**4.1 The 1:9 MMR with Venus**

We found 720 $m_1$ particles, 699 $m_2$ particles, and 674 $m_3$ particles. All of them did not initially eject onto near-resonant orbits. Let us consider the most numerous $m_1$ particles. Figure 4 plots the typical behavior of these particles. They are observed in the near-resonance region and until the year 3000, a large amplitude with irregular oscillations is observed near the resonant gap α. Its argument β circulates mostly with few efforts to transfer to libration at small periods not exceeding 40 or 50 years, thereby confirming its location near the 1:9 resonance with Venus. This is supported by the behavior of the resonant gap passing through zero at abovementioned periods; the argument circulates beyond these periods without deviations for about 70 years. The semi-major axis periodically changes during the same period (see Fig. 2 case I). The 1:9 MMR with Venus is unstable during the whole period, and the particles are close to the resonance region. The similar behavior is observed for meteoroids with masses $m_2$ and $m_3$.

TABLE 1. Resonant Particle Number on 3000 Modeled Meteoroids in the Quadrantid Stream

| MMR | Masses of asteroid | | |
|---|---|---|---|
| | $m_1$ | $m_2$ | $m_3$ |
| 1:9V | 728 | 699 | 674 |
| 1:5E | 83 | 80 | 75 |
| 1:3M | 577 | 560 | 594 |
| 3:8M | 72 | 63 | 58 |
| 2:1J | 2 | 0 | 9 |
| 7:3J | 96 | 92 | 92 |
| 9:4J | 359 | 394 | 369 |

**4.2 The 1:5 MMR with Earth**

We found a small number of particles captured by the 1:5 resonance with Earth, i.e., 83 $m_1$ particles, 80 $m_2$ particles, and 75 $m_3$ particles. These particles (~8%) behaved similarly to the resonance, as presented in Fig. 5. In the upper diagrams, the particle was close to the resonance region, although its interaction with the 1:5 resonance with Earth was slightly weaker than that shown in the bottom diagrams. For $m_1$ particles (Fig. 5, upper diagrams), in the period from the ejection till the year 2150 and from 2750 till 4000, the libration center of the resonant gap α with a small oscillation amplitude, shifted toward negative values, while during entering the resonance, oscillations had the same periods relative to 0 and the amplitude of 30″/day. Circulation without resonance transferred to a slow circulation, and after the year 3460, it transferred to libration with the center shift. $m_1$ particles closely approached to the Earth on 31.12.1845 (0.030 a.u.), 31.12.2181 (0.097 a.u.), and 28.06.3460 (0.09 a.u.), that, in turn, led to a chaotic particle motion, and after close approaches it was unpredictable.

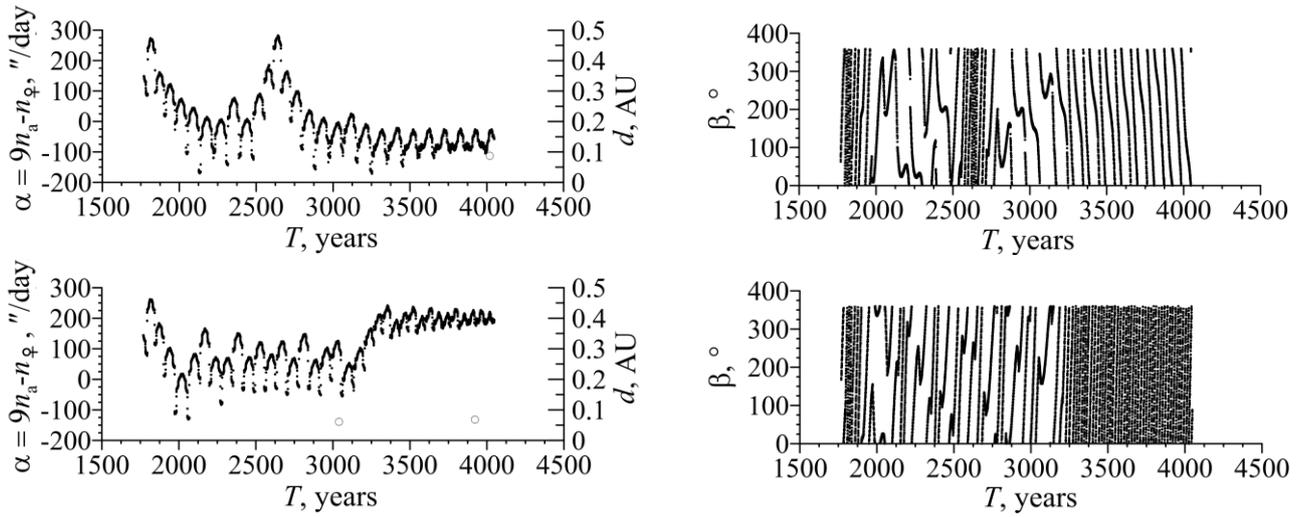

Fig. 4. The 1:9 MMR with Venus: resonant gap α with overlapped approaches (left) and critical argument β (right) for two $m_1$ test particles.

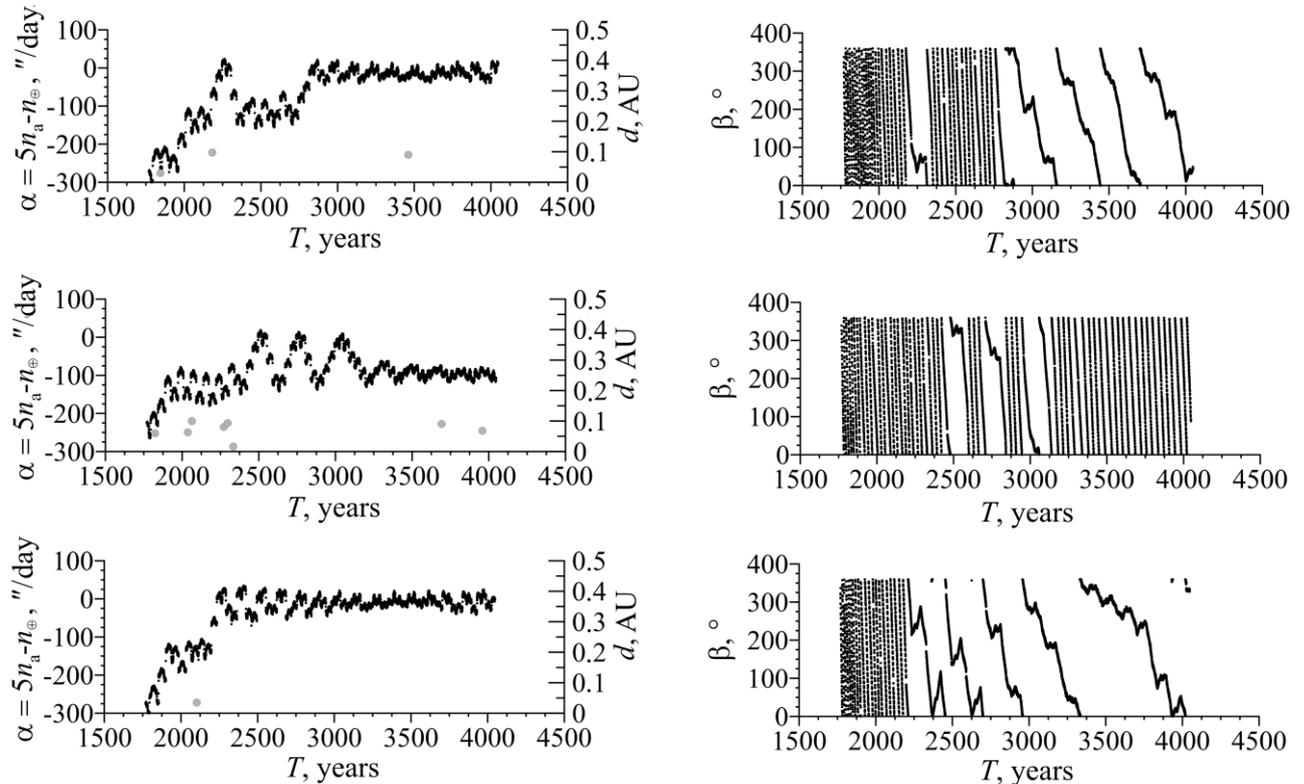

Fig. 5. The 1:5 MMR with Earth: resonant gap α with overlapped approaches (left) and critical argument β (right) for three $m_1$, $m_2$, $m_3$ test particles.

In the case with $m_2$ particle (e.g., Fig. 2, case III), one can see several capture efforts just at small periods, but resonance is absent within the whole period. This unstable geometric configuration particle–Earth results in close approaches to the Earth on 2.1.1823 (0.059 a.u.), 1.1.2037 (0.062 a.u.), 8.1.2063 (0.09 a.u.), 6.1.2271 (0.08 a.e.), 4.1.2297 (0.09 a.u.), 6.1.2333 (0.016 a.u.), 7.4.3691 (0.09 a.u.), and 27.4.3958 (0.067 a.u.) (Fig. 5, middle diagrams). This leads to changes in the semi-major axis and the creation of mean motion commensurability.

In the case with $m_3$ particle (e.g., Fig. 2, case IV), the situation is similar to that with $m_1$ particle, but oscillations occur after the year 3000 with the shift of the libration center at an amplitude of 50″/day.

### 4.3 The 1:3 and 3:8 MMR with Mars

The 1:3 resonance with Mars is similar to the 1:9 resonance with Venus, and just over half of test particles undergo this resonance (see Table 1). For example, $m_2$ particle in Fig. 6, tries to remain in the resonance region in the period from 2300 till 2600, but librations of argument β are not regular. Note that they have regular circulations of the argument β till 2300, after which transfer to libration with the shifted center.

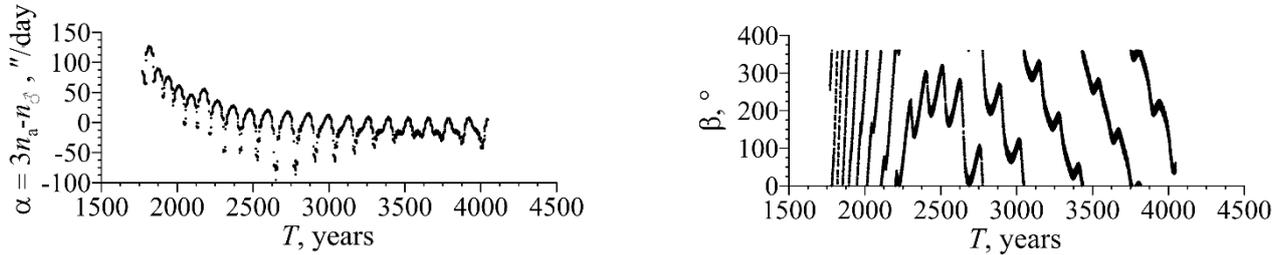

Fig. 6. The 1:3 MMR with Mars: resonant gap α (left) and critical argument β (right) for $m_2$ test particle.

We have found small number of particles captured by the 3:8 resonance with Mars (Fig. 7). Librations of argument β are not regular in this case. Unfortunately, discovered particles are not always in this resonance, just at small periods after the year 2500 they are (Fig. 7, upper diagrams). In these diagrams, librations of argument β are observed for the test particle $m_3$ in the period from 2600 till 3200. The resonant ratio passes through the zero crossing, and the libration/circulation of the resonant argument constantly shifts to the libration center. The test particle $m_3$ is captured by the resonance region, amplitudes of the resonant gap α and the argument β reduce, the object occurs in the resonance in the period from 3000 till 3700, and then leaves it again.

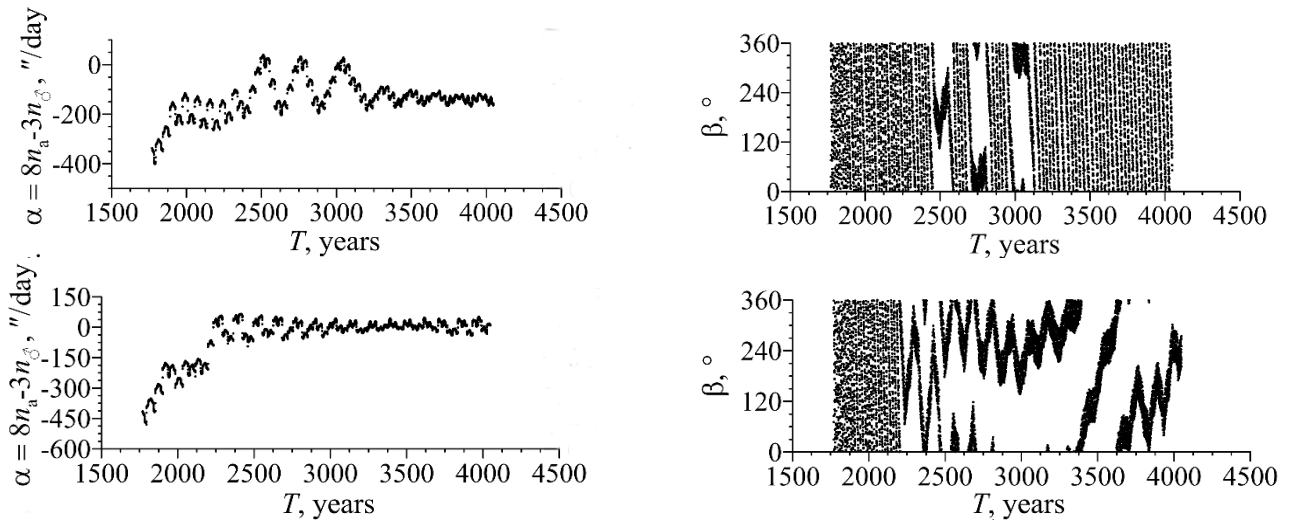

Fig. 7. The 3:8 MMR with Mars: resonant gap α (left) and critical argument β (right) for two test particles $m_2$ and $m_3$ (top-bottom).

### 4.4 THE 2:1, 7:3 AND 9:4 MMR WITH JUPITER

Although the correlation between the Quadrantid stream and the 2:1 resonance with Jupiter manifests itself slowly and is complicated, the proximity of this resonance even in nonresonance state strongly affects the motion of model meteoroids, which is described in detail in our previous research [7]. The semi-major axis $a$ demonstrates periodical changes with period of about 60 years (see Figs 1, 2, 8), which is also observed in the simulation of the Quadrantid stream [4]. Near-resonant state can have an additional effect on this stream, since particles ejected from 2003 EH1, are more likely to be in this resonance. Precession rates of orbits of meteoroids trapped in the resonance, notably differ from those presenting beyond the resonance.

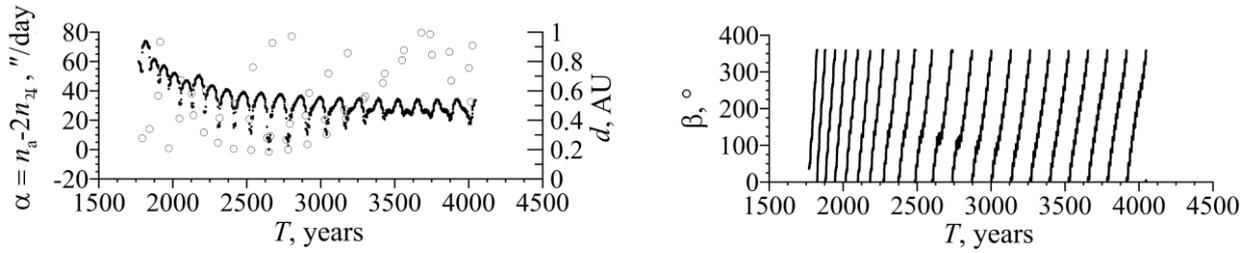

Fig. 8. The 2:1 MMR with Jupiter: resonant gap α with overlapped approaches (left) and critical argument β (right) for $m_2$ particle.

We discover 359 particles $m_1$, 394 particles $m_2$, and 369 particles $m_3$ trapped in the 9:4J resonance in the far future, and 96 particles $m_1$, 92 particles $m_2$, and 92 particles $m_3$ partially trapped in the 7:3J resonance. All of these particles have not been initially released onto resonant orbits and, as shown in Fig. 9, some of them are in 7:3J and 9:4J resonances in the far future. The motion of model meteoroids is strongly affected by the 2:1J resonance and their periodical trapping by 7:3J and 9:4J resonances for short periods of time. Investigation of the evolution of the nominal orbit and the orbit of the test particle (see Fig. 2), shows that orbital elements are exposed to a series of low-amplitude periodic oscillations. Orbital elements, the semi-major axis in particular, suffer rapid and great changes induced by closest approaches to Jupiter. High MEGNO values (Fig. 2) are associated with frequent changes in the semi-major axis caused by multiple approaches to Jupiter nearby the Hill sphere. For example, in Fig. 2 (cases I, III IV), the released particle equals the semi-major axis value after the year 1780, which differs from the former by ~0.2 a.u. of the value after the year 2600, as Jupiter induces significant oscillations when the particle was in 7:3J and 9:4J resonances. Sometimes, the critical argument circulates and sometimes it librates at a high amplitude, namely: from 1780 till 2020 for the case III (Fig. 9, upper diagrams) and from 2500 till 3000 for the case IV (Fig. 9, middle and bottom diagrams). It can be thus concluded that the particles are nearby resonances, but not captured by them. According to [17–19], the MMR of the motion is unstable and can cause a chaotic motion. In [17, 18], it is shown that overlapping of different resonances can result in chaotic motion of small bodies of the solar system, especially in the case with at least one unstable resonance, i.e., resonant argument changes its libration into circulation and *vice versa*. The same behavior is observed for $m_1$ particles.

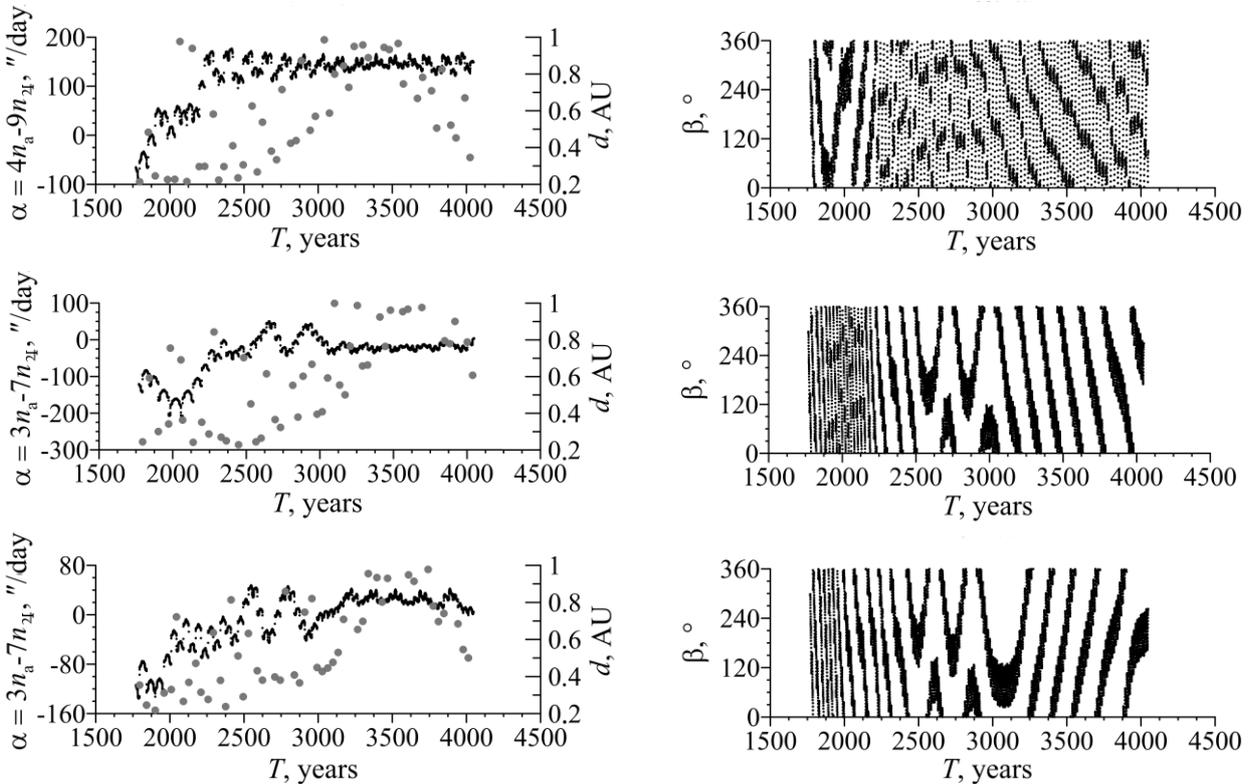

Fig. 9. Resonant gap α evolution with overlapped approaches to Jupiter (left) and critical argument β (right) for the 9:4J resonance of test particle $m_2$ (upper diagrams), 7:3J resonance of test particle $m_3$ (middle and bottom diagrams).

## CONCLUSIONS

In this work, we have presented one of the possible scenarios for the dynamical evolution of the meteoroid stream formed by the asteroid (196256) 2003 EH1. The dynamics analysis of meteoroid particles identified a complex dynamic structure of the stream. Investigations concerned the orbit space of the simulated Quadrantid meteoroid stream consisting of particles with mass 0.0003, 0.03 and 0.3 g. The numerical calculation of the meteoroid behavior showed that they could be in six mean-motion resonances for about 3300 years, namely: in the 1:9 resonance with Venus, in the 1:5 resonance with Earth, in 1:3 and 3:8 resonances with Mars, and in 7:3 and 9:4 resonance with Jupiter. High MEGNO values were associated with frequent changes in the semi-major axis caused by multiple approaches to planet resulting in the formation of the mean motion commensurability. Some of test particles were in the vicinity of these MMR for some periods, that affected their evolution. They either moved nearby the resonance region, or enter or leave it.

Research findings were obtained for the model, which did not completely match the real Quadrantids' location and was used for the simulation only. These results supposed more research into the observed Quadrantids, but resonant traces might be far from the Earth's orbit.